# Colors Of Graphite On Silicon Dioxide

S. Roddaro,[1,*] P. Pingue,[1] V. Piazza,[1] V. Pellegrini,[1] and F. Beltram[1]

[1]*NEST-INFM & Scuola Normale Superiore, Piazza dei Cavalieri 7, I-56126 Pisa, Italy*

Monoatomic layers of graphite can be electrically contacted and used as building blocks for new promising devices. These experiment are today possible thanks to the fact that very thin graphite can be identified on a dielectric substrate using a simple optical microscope. We investigate the mechanism behind the strong visibility of graphite and we discuss the importance of the substrate and of the microcope objective used for the imaging.

PACS numbers: to be added

Thin graphite has recently attracted a large interest in the scientific community since it was shown that single layers of graphite can be individually contacted and used as device building blocks [1–6]. A graphite monolayer, also known as *graphene*, indeed constitutes a quite unique case of stand-alone two-dimensional electron system with remarkable properties: its mobility is very good even at room temperature and both $p$-type or $n$-type conduction can be easily induced by external gating [1, 2, 6]. In addition, graphene band structure displays a set of very unusual features that lead to a set of interesting new phenomena in transport such as the anomalous integer quantum Hall effects [3–5].

Graphene samples are typically obtained by a simple method of mechanical exfoliation on a $SiO_2/Si$ substrate [1, 2]. This technique yields random flakes of many different thicknesses and for this reason the identification of very thin graphite on $SiO_2/Si$ is a crucial step in the fabrication of graphene-based devices. The critical issue of finding the "good" graphite flakes has so far been solved in first place by using simple optical microscopy. This technique still remains a compulsory fabrication step as it allows a quick thickness survey before more precise but less direct methods such as Raman spectroscopy [7, 8] are used to selectively investigate small portions of the sample. Despite this importance, optical evidences of thin graphite reported in literature still appear to be controversial: in some cases single monolayers are said to be completely *invisible* [1, 2] while other experiments imply that graphene can actually be seen optically in a quite clear way [5]. The same interpretation of the effect still does not appear to be well-established within the graphene community [8]. In this Letter we focus on the physics behind the strong visibility of thin graphite on $SiO_2/Si$ substrates and we present a simple model that reproduces quite well the observed graphite colors. Our analysis indicates that the most cited explanation, based on the phase-shift in the interference color, is not catching the main mechanism behind the effect. We discuss the importance of the substrate and the objective used in the determination of graphite visibility.

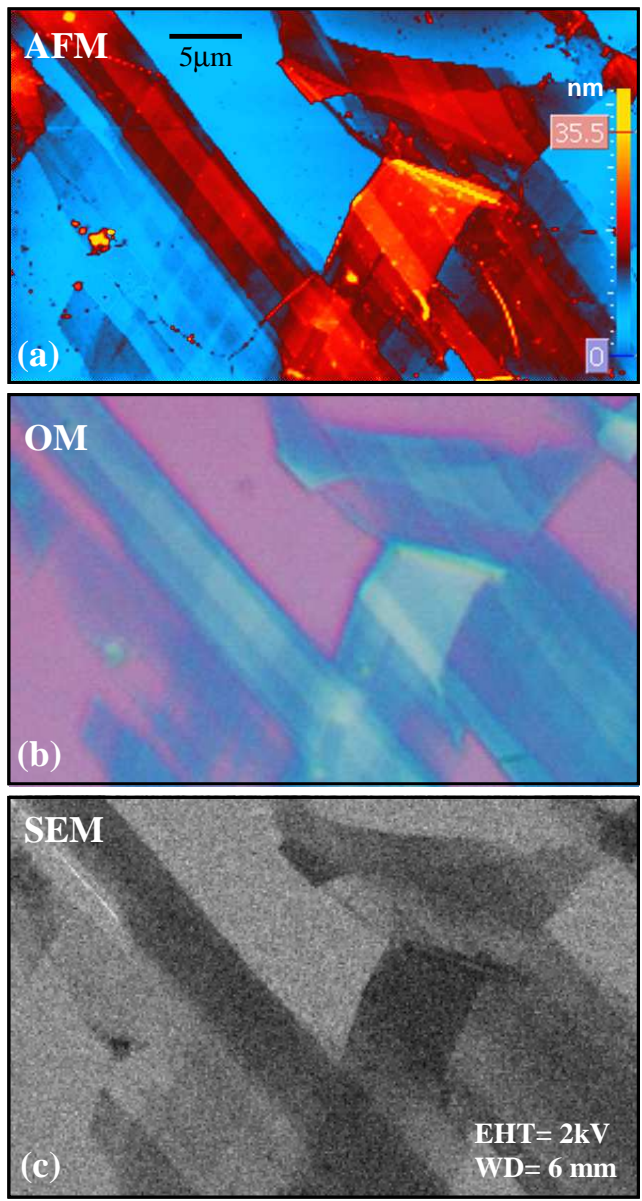

FIG. 1: Atomic force image (a), optical image (b) and scanning-electron microscope image (c) of a big graphite flake containing regions with many different thicknesses.

---

*Electronic address: `s.roddaro@sns.it`

We propose possible directions of optimization for this optical imaging technique.

An example of graphite apparent colors on SiO$_2$/Si is reported in Fig. 1a for a substrate with an oxide thickness $\Delta_{SiO_2} = 500$ nm. The optical microscope (OM) picture of Fig. 1a is compared in Fig. 1b with a topographic image obtained using an atomic force microscope (AFM). Regions with a thickness of few monolayers appear to be barely visible optically on Fig. 1a. Even if AFM yields precise information about graphite thickness, it cannot be used to probe systematically the graphite thickness over a big sample. Figure 1c shows an additional picture taken using a scanning electron microscope (SEM) at low acceleration voltage. Visibility of thin graphite appears to be slightly better then in OM but, on the other hand, SEM imaging can be a rather risky technique as it tends to deposit amorphous carbon on the observed samples. In conclusion, OM imaging is at present time the only technique that can be used extensively and conveniently over a macroscopic sample. Once good graphite flakes have been spotted, different options exist for a precise but spatially limited thickness evaluation. Beside AFM, one interesting case is represented by Raman spectroscopy [7, 8], which appears to provide a conclusive thickness evaluation. Another important test consists in the verification the anomalous transport behavior expected for single graphene layers [3, 4]. Issues related to inter-layer coupling in real graphite [9], however, cast additional doubts on the estimated thickness of the measured sample.

It is well-known that the thickness of a SiO$_2$ film grown on top of a Si substrate can be evaluated with some precision simply by *looking at it* [10]. The apparent color of such a substrate is in fact due to the interference between the different reflection paths that originate from the two interfaces air-to-SiO$_2$ and SiO$_2$-to-Si. Depending on the distance between these, the various interfering paths will experience relative phase shifts and thus thickness variations of a fraction of wavelength will give color shifts that can be easily appreciated just by a quick look. Even when considering this interference effect, the visibility of thin graphite on SiO$_2$/Si still appears to be striking as it is generally agreed that few monolayers, with a total thickness of the order of 1 nm (the distance between crystalline planes in graphite being about 3.4 Å), can still be identified [1, 2]. This thickness accounts for barely one or two parts over 1000 of the average wavelength in air for visible light: a remarkably small fraction to be appreciated by eye, even when phase contrast is involved. Indeed, here we show that a simple interpretation in terms of phase shifts is incorrect. Most of the effect is in fact due a modulation of the *relative amplitude* of the interfering paths as a consequence of the simple fact that graphite, being a conductor, has a transparency that depends on thickness in a sensitive way. Relative amplitude modulations are then made strongly visible by a fortuitous combination of permittivities in the SiO$_2$/Si multilayer. This leads to a resonant cancellation of reflection by destructive in-terference for specific wavelengths and for a finely tuned and relatively small thickness of graphite.

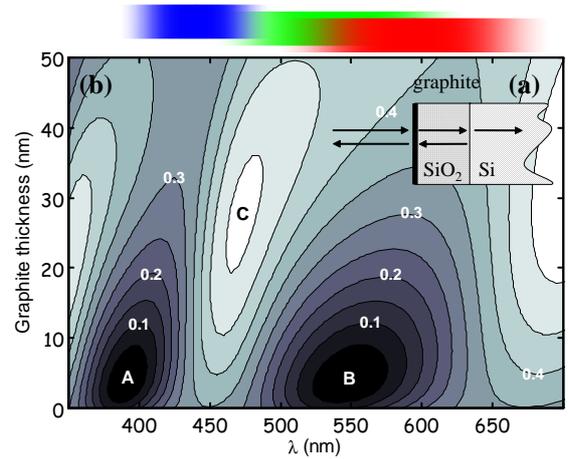

FIG. 2: (a) Multilayer structure used in the transfer-matrix simulation. (b) Calculated reflection spectrum for a 460 nm-thick SiO$_2$/Si substrate and a variable graphite thickness $\Delta_G$. Graphite *enhances* the modulation in the reflection spectrum which is already present at $\Delta_G = 0$ and is due to interference in the SiO$_2$/Si substrate. This leads to a set of finely tuned conditions for complete suppression of reflection for $\Delta_G \approx 5$ nm (minima **A** and **B**). This is the major reason for thin graphite visibility. Spectrum also displays a red-shift due to increased length of the optical path through the graphite. For $\Delta_G > 50$ nm oscillations in the reflection spectrum tend to disappear due to skin effect [1, 2]. On top we report the approximate red, green and blue sampling regions for a standard trichromatic device.

In order to explore more in details this mechanism, we report the results of a very simple bulk model for the graphite/SiO$_2$/Si multilayer (see Fig. 2a). Our calculation is based on classical electrodynamics and on the transfer matrix formalism for the evaluation of the reflection spectrum at normal incidence [11]. As the calculation is standard and easily reproducible, we will not go into its details and we will rather focus on our modeling of the materials involved. The layers were described using complex relative dielectric constants $\epsilon_r = \epsilon_1 + i\epsilon_2$. For air, silicon and SiO$_2$ we adopted, following standard refractive index values [10]. Graphite was described by $\epsilon_r \approx 4.5 + 7.5i$, following experimental data in the optical region of the spectrum [12, 13]. This value yields a skin depth which is consistent with the observed 50 nm opacity crossover [1, 2]. Despite our rather approximated description of the multilayer, we show that the model reproduces quite clearly the mechanism behind the evolution of the apparent color of graphite as a function of the thickness.

Figure 2b shows the contour plot of the reflection spectrum $r(\lambda, \Delta_G)$ as a function of both the wavelength $\lambda = 2\pi c/\omega$ and of the thickness $\Delta_G$ of the graphite. The reflection values were calculated for $\Delta_{SiO_2} = 460$ nm: the reason behind this choice will be discussed later in the paper. The plot shows rather clearly that the presence of thin graphite on the surface indeed produces a red shift of the interference in the reflection spectrum, as a consequence of the additional optical length introduced in the system. However Fig. 2b also highlights that in the region of $\Delta_G < 10$ nm: graphite layers *enhance* the interference effects already present at $\Delta_G = 0$ and lead to a resonant suppression of the reflection for specific matching conditions in the $(\lambda, \Delta_G)$ parameter space. In the specific case reported here we get reflection zeros for $\lambda \approx 390$ nm and $\lambda \approx 550$ nm within the visible spectrum, in both cases the zeros occur at a graphite thickness $\Delta_G \approx 5$ nm. This effect is due to a *modulation* of the transparency of the air-graphite-$SiO_2$ interface which allows light to enter in the $SiO_2$ layer: only for a finely tuned thickness $\Delta_G$ it is possible to meet the conditions for a completely destructive interference in the reflection. The whole system can also be thought as a sort of Fabry-Perot cavity where graphite plays a role in balancing the input/output barriers and thus in maximizing the interference contrast. We argue that the presence of these reflection zeros is the main reason for the strong visibility of thin graphite on $SiO_2$.

The model used here is rather simple and adopts a bulk approximation for the layers. Despite the rude approximations, our results match closely the observed color evolution of a real graphite flake, as we show in the following. In order to make this clear, we approximately reproduce the visible colors corresponding to $r(\lambda, \Delta_G)$ using the standard RGB trichromatic coding for display devices. We calculate the RGB values corresponding to the light spectrum $r(\lambda, \Delta_G)$ following

$$C_i(\Delta_G) = \int r(\lambda, \Delta_G) f_i(\lambda) d\lambda, \quad i = R, G, B \quad (1)$$

where we use the color matching functions $f_i(\lambda)$ defined by the CIE standards [14] and $C_i(\Delta_G)$ represents the $i$ component in the RGB color coding for a specific $\Delta_G$.

Figure 3 shows the calculated color vs. thickness calibration and compares it with normal interference colors for $SiO_2$. The color scale of Fig. 3b span a thickness of 80 nm and the 5 nm on the left side of the 0 nm point reports the original color of $SiO_2$. Figure 3c shows a smaller thickness range up to 10 nm. Our calculated color bar matches quite well the real colors recorded by a camera in a normal microscope set-up (Fig. 1b and 3a). Figure 3d reports the colors obtained for a simple accumulation of phase shift due to the longer interfering paths: the color evolution is similar (at least up to 10 nm) but the contrast for small thicknesses results to be much reduced. Many of the features of the $r(\lambda, \Delta_G)$ reported in Fig. 2 can be quickly seen just by comparing the AFM picture of Fig. 4a with the RBG components of Fig. 3a, which are reported in Fig. 4b,c and d. Taking AFM data as a thickness reference, it is clear that when $\Delta_G$ increases, re-flection in the red channel first decreases gradually, then reaches a broad minimum for a thickness in the range $\Delta_G \approx 5 - 15$ nm and finally increases again. This behavior is consistent with results reported in Fig. 2 as the red component will just sample the region of spectrum interested by the reflection minimum **B**. The green channel in Fig. 4c shows a similar evolution even if this component appears to give more contrast on thicker graphite.

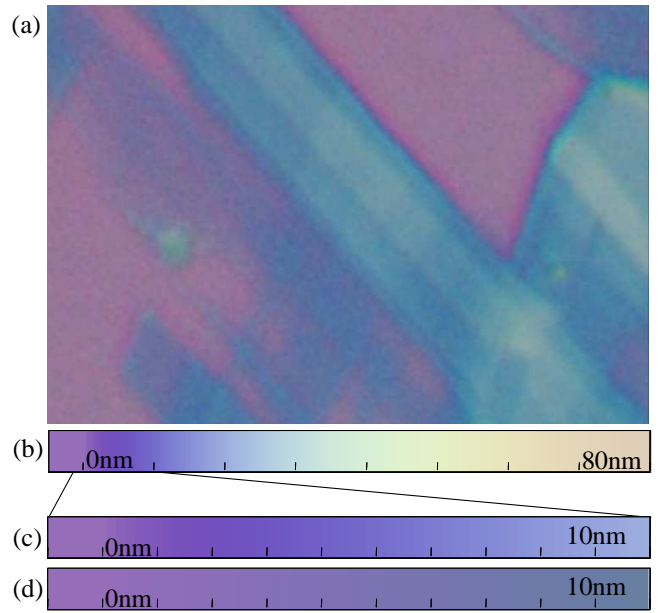

FIG. 3: An optical image of a graphite flake (a) is compared with our simulation of the graphite color as a function of the thickness (b and c). Color evolution due to simple phase shift in the interference paths (d) would give a much reduced contrast on small thicknesses.

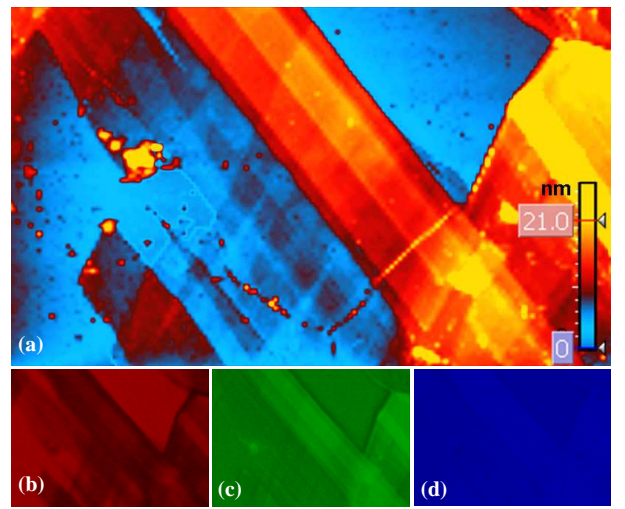

FIG. 4: (a) AFM image of the graphite flake depicted in Fig. 3a. (b), (c) and (d) RGB components of the picture reported in Fig. 3a.

Again this is consistent with what expected from Fig. 2: the green sampling region is still marginally interested by the minimum **B** but it is also affected by the maximum **C** giving enhanced reflection for thicker graphite. Finally, Fig. 4d finally shows that a very small color variation is detected on the blue component of the image, because in this case none of the regions of strong $\Delta_G$-dependent modulation in the reflection is sampled very well.

Having showed here that we can reproduce the experimentally observed graphite colors, we now focus on the intrinsic limits of thickness evaluation by OM and provide a few guidelines for improving this technique. The main problem with OM is that incidence and reflection of light on the sample is *not* orthogonal: optical beams in a microscope are in fact conical, as quantified by the numerical aperture (NA) of the objective. Light hits graphite and $SiO_2$ layers at various angles simultaneously and both the length of the interfering paths and the reflection coefficient of the various interfaces will be affected. As a consequence, different matching conditions $(\lambda, \Delta_G)$ will be obtained for the reflection cancellation of Fig. 2 depending on the incidence angles and polarization of light. Due to the rather poor coherence of light sources in OM, we expect that the light collected at different angles will sum up in an incoherent way inside the microscope optical system. Features in Fig. 2 will thus result to be *smoothed-out* and the visibility of thin graphite on $SiO_2$ will eventually decrease. The importance of such an effect can be quantified by taking a standard NA $\approx 0.7$. A simple calculation shows that the corresponding deviation angle $\theta$ in the $SiO_2$ will be at most about 20 degrees. We do not report here detailed calculations but we note that if one neglects the small angle-dependent variations in the reflectivity of the interfaces, the multilayers will just behave as if the effective $SiO_2$ thickness were $\Delta_{SiO_2,\text{eff}} = \Delta_{SiO_2} \cos\theta$. This equation and our NA are consistent with the fact that we can reproduce the graphite colors using a value of $\Delta_{SiO_2} = 460$ nm, i.e. approximately 10% smaller than the nominal value for our wafer (500 nm). Similar calculations performed at $\Delta_{SiO_2} = 230$ nm and $\Delta_{SiO_2} = 270$ nm can reproduce graphite apparent colors observed on substrates with 250 nm-thick (see supplementary material) and 300 nm-thick [2] oxide layers.

The analysis above suggests possible directions for improving this technique. For instance we expect that a smaller NA will increase the contrast of thin graphite even though the image details will probably be affected in a negative way. An improvement should be observed when using more refractive dielectric media such as $Si_3N_4$ as this will obviously decrease the incidence angle in the dielectric for a given value of the NA. In addition, a dielectric with different refraction index will also lead to a different critical graphite thickness $\Delta_G$ at the reflection zeros and this will influence the contrast of thin graphite. We also point out that having a higher light coherence, in particular in the regions of the spectrum interested by the reflection minima, is also likely to be useful in reducing the incoherent averaging linked to a large numerical aperture. We finally note that, as long as direct eye observation of graphite is involved, the details of eye sensitivity in the visible spectrum will probably lead to better or worse thickness evaluations depending on the substrate, which determines the frequency position of the reflection minima. We believe however that it will be difficult to reach reliable and well-established conclusions on this issue as long as all the intrinsic limitations of the OM method, in particular those linked to the value of the used NA and thus to the presence of non-perpendicular incidence, will not be kept under appropriate control.

In conclusion, we discussed the electrodynamics behind the visibility of few monolayers of graphite deposited on thin $SiO_2$ and we showed that the visibility of thin graphite is linked to a strong amplitude modulation of reflection at the air-graphite-$SiO_2$ interface, while the modulation of the optical lengths appears to play a marginal role. Finally we note that a simple extrapolation from our model yields a reduction of reflection between $\Delta_G = 0$ and $\Delta_G = 0.34$ nm (graphene's thickness) in excess of 10% in the spectral regions corresponding to the minima in Fig. 2. In principle, once the issues of perpendicular incidence are solved, such a color modulation should be easily detectable and the effect could be used for a reliable thickness evaluation of graphite.